# Supercontinua for high resolution absorption multiplex infrared spectroscopy

Julien Mandon [1], Evgeni Sorokin [2], Irina T. Sorokina [3], Guy Guelachvili [1], Nathalie Picqué [1]

[1] Laboratoire de Photophysique Moléculaire, CNRS, Université Paris-Sud, Bâtiment 350, 91405 Orsay Cedex, France
[2] Institut für Photonik, TU Wien, Gusshausstrasse 27/387, A-1040 Vienna, Austria
[3] Norwegian University of Science and Technology, Department of Physics, N-7491 Trondheim, Norway

**Abstract:** Supercontinua generated in highly non-linear fibers by ultrashort-pulse lasers can be used for high resolution Fourier transform absorption spectroscopy. The practical advantages of these bright ultrabroadband light sources for spectroscopy are reported in the near-infrared region. A $Cr^{4+}$:YAG femtosecond laser broadened by an extruded soft-glass photonic crystal fiber, emitting from 1200 to 2200 nm and from 675 to 950 nm, provides a spectral radiance being $1 \cdot 10^5$ times higher than that of a 3000 K blackbody and $10^2$ times higher than that of a synchrotron radiation. The $C_2H_2$ and $NH_3$ overtone spectra are recorded using this source within a few seconds.

OCIS codes: 190.4370 Nonlinear optics, fibers; 060.2390 Fiber optics, infrared; 140.7090 Ultrafast lasers; 230.6080 Optical devices, Sources; 300.6390 Spectroscopy, molecular; 140.3070 Infrared and far-infrared lasers

Bright, coherent and broadband radiation brings a number of advantages for high resolution spectroscopy, so that a number of several facilities are, for instance, presently using [1] - or planning to use - synchrotron radiation for gas phase spectroscopy. As a much simpler table-top low-cost alternative, femtosecond (fs) mode-locked lasers represent the first versatile coherent sources providing broad bandwidth and high brightness. Several spectroscopic methods using fs mode-locked lasers are currently being implemented. Most of them make use of multichannel grating spectrometers [2,3], which limit the resolution and applicable spectral domains. Multiplex spectroscopy does not meet these restrictions and various approaches have been successfully demonstrated [4-6], in particular in the infrared where detector arrays for multichannel recording are expensive or inexistent. These approaches are rather easy and cheap to be put in practice with commercial and laboratory instruments widely used in chemistry and physics. However, their full potential is currently limited by the narrow spectral range of available fs sources in the infrared. The supercontinuum (SC) generation may be a convenient and inexpensive way to overcome this restriction.

SC can be conveniently generated from femtosecond lasers or any high energy pulsed lasers when launched into nonlinear optical fibers. In particular, Photonic Crystal Fibers (PCF) have permitted more than one octave spectral broadening in the near-infrared and visible [7]. Currently, SC is boosting several fields like frequency metrology, pulse compression and optical coherence tomography. Its application to optical gas monitoring is presently restricted to wavelength agile methods [8-10]. Multiplex Fourier Transform (FT) spectroscopy is an interesting opportunity to benefit from these broad sources for high accuracy and high sensitivity gas phase spectroscopy. The aim of this paper is to show that SC may be fully taken advantage of in multiplex spectroscopy and to illustrate the SC potential for high resolution FT spectroscopy on the example of a SC generated in a highly nonlinear fiber coupled to a 1.5 µm fs laser.

The experimental setup is shown in Fig 1. A chirped-mirror-controlled $Cr^{4+}$:YAG oscillator [11], pumped by a $Nd:YVO_4$ laser, is modelocked by a commercial SESAM (BATOP Mod. SAM-1550-1). It is operated in the chirped-pulse positive dispersion regime [12]. In this regime, details of which are described elsewhere [13], the laser emits positively chirped 1.4 ps pulses centered at 1525 nm with a repetition rate of 142 MHz and 100 mW average power. These pulses are compressed down to 120 fs in a 3.2 m piece of a standard SMF28 single mode silica fiber, and launched into the nonlinear fiber with about 550 pJ input energy. We have tested several kinds of nonlinear fibers, including dispersion-shifted germanium-doped silica fibers (DSF) and PCFs. The most effective broadening has been obtained with a 40-cm long PCF made of SF6 glass [14] in which the light, launched with an efficiency of about 32%, is guided through a 4.5 µm-diameter solid core with nonlinear index $n_2 = 2.2 \times 10^{-15}$ $cm^2.W^{-1}$, surrounded by a periodic arrangement of air holes. After recollimation, the SC beam has about 25 mW average power and passes through a single-pass 70-cm-long absorption sample cell. The beam is attenuated and analyzed by a commercial rapid-scan FT spectrometer (Bruker IFS66) just as a steady-state source.

The achieved spectral broadening is shown on Fig. 2 at low resolution (50 $cm^{-1}$ - ca. 10 nm). The spectral domain of the laser emission itself (without broadening) extends over 200 nm at -30 dB level. In the high resolution spectra reported below, this 30 dB attenuation corresponds to the limit beyond which no spectroscopic data are measurable. At the PCF output, the spectrum extends in the infrared over about 1100 nm, from 1200 (8330 $cm^{-1}$) to 2200 nm (4545 $cm^{-1}$) at -30 dB level, and in the visible-very near infrared over about 275 nm, from 675 (14815 $cm^{-1}$) to 950 nm (10525 $cm^{-1}$) at -30 dB level. This broadening is consistent with previous experimental data [14-17] on the same type of fiber with different core sizes and various seeding sources.

Each high resolution (3.6 GHz - 0.12 cm$^{-1}$) spectrum of the PCF output, is recorded in in just 13.1 s. It is made of 67290 independent useful spectral elements covering 1375 nm (8075 cm$^{-1}$), with a maximum signal to noise ratio (SNR) about 2400. At this resolution, fringes scrambling spectral information are observed. Their origin is not completely clear: they do not seem to be due to birefringence from the PCF as rotating the input light polarization has no effect. They are however specific to the employed PCF, because the SC generated in a single mode DSF is fringe-free. We believe that these fringes are due to the mode beating in the multimode PCF that could not be observed previously due to insufficient resolution [16]. The overall short-term stability made possible the efficient suppression of the fringes, by dividing the spectra with and without gas inside the absorption cell. Since the broadening was much stronger in the multimode PCF, we decided to use it to test the SC potential for high resolution spectroscopy, trading the slightly degraded SNR against the larger bandwidth.

As a first demonstration, $C_2H_2$ and $NH_3$ spectra are recorded. Figure 3 shows a portion of the acetylene absorption spectrum at two different pressures: 28 and 160 hPa, between 1511 and 1548 nm (6620-6460 cm$^{-1}$) at 3.6 GHz (0.12 cm$^{-1}$) resolution. All absorption lines displayed in Fig. 3 are due to acetylene. The most intense band is the $\nu_1+\nu_3$ band of $^{12}C_2H_2$. The rotational lines from the P and R branches of the $\nu_1+\nu_3+\nu_4^1-\nu_4^1$ and $\nu_1+\nu_3+\nu_5^1-\nu_5^1$ hot bands of $^{12}C_2H_2$ and the $\nu_1+\nu_3$ cold band of $^{12}C^{13}CH_2$ are also observed. The ammonia spectrum is shown on the lower trace of Fig.4 at 15 GHz (0.50 cm$^{-1}$) resolution. The region of the $\nu_2+\nu_3+\nu_4$ band of $NH_3$ is observed. The spectrum is complex and crowded and to our knowledge, no rotational assignment is available in the literature. For comparison, the upper trace provides a portion of spectrum recorded with a tungsten lamp, which is the most widely used white source in near-infrared absorption FTS. The SNR of this reference spectrum is about 1.7 times better than the SC spectrum, but the recording time is 200 times longer. With the same recording time as with the tungsten lamp, the SNR of the SC spectrum would have been 33 000. One advantage of SC sources is their high spectral radiance, which for the present 25 mW SC source equals $2.8 \times 10^7$ W.m$^{-2}$.sr$^{-1}$.nm$^{-1}$. It is $1 \times 10^5$ times stronger than a tungsten lamp comparable to a 3000 K black body source. A SNR enhancement of about 350 could have then been achieved. This was actually not possible: the detector was saturated by the 25 mW laser beam and a power attenuation by more than one order of magnitude before entering in the interferometer was necessary. On the other hand, tungsten lamps have a flat spectrum, which enables to benefit from the whole spectrum. Further optimization of SC generation providing flat spectra without strong intensity variations would improve their applicability for spectroscopic measurements.

It is also informative to compare the potential of SC sources with the performance of synchrotron radiation. The expected brightness [18] of the infrared beamline at the French facility Soleil is $10^{19}$ photons/s/1% of BW/mm$^2$.steradian for the near infrared (1 µm), which converts into a spectral radiance equal to $2 \times 10^5$ W.m$^{-2}$.sr$^{-1}$.nm$^{-1}$. This is two orders of magnitude weaker than the present SC, which can still be improved in terms of output power and spectral coverage. Moreover, since such SC source is a table-top device made of commercially available parts, it could even be incorporated into industrial instruments.
Presently the wavelength agile techniques are the fastest gas monitoring methods using the SC sources at comparable resolution [8-10]. They, however, feature frequency-dependent spectral resolution and require long fibers as dispersers, that currently limit their applicability to the telecom bands and have wavelength sweeps that may be too fast. A FT spectrometer requires longer acquisition times, seconds instead of milliseconds, but is much more appropriate for accurate measurements of spectroscopic interest. The FT spectral resolution may be improved easily up to 30 MHz, the accuracy in the wavenumber scale may reach $10^{-9}$, and several millions of spectral elements may be measured in a single experiment, resulting in spectra spanning more than one octave. In addition, SC can be implemented with the frequency comb spectrometric approach that we reported recently with a mode-locked laser [6], which enhances sensitivity and enables simultaneous recording of broadband absorption and dispersion spectra.

In this article, supercontinuum femtosecond laser systems are demonstrated to be powerful high-brightness sources for high resolution spectroscopy. The corresponding accurate analysis may be well performed by FT spectrometers over the whole emission range. SC sources become commercially available in the visible and near-infrared ranges and they may be easily implemented on the basis of existing spectroscopic instruments. The development of a table-top ultrabroadband coherent source in the mid/far-infrared should significantly impact the practical applications of spectroscopy, and SC generated by chalcogenide or fluoride fibers [19] could likely be an efficient approach.

O. Okhotnikov (Tampere University of Technology) is warmly thanked for providing the samples of nonlinear Ge-doped fiber, P. Jacquet (Laboratoire de Photophysique Moléculaire, Orsay) for his participation and M. Hanna (Laboratoire Charles Fabry de l'Institut d'Optique) for useful advices. Efficient technical support was provided by J. Caillaud. I.T Sorokina acknowledges an Invited Professor position with Université Paris-Sud. The Programme Pluri-Formation de l'Université Paris-Sud "Détection de traces de gaz" and the Austrian Fonds zur Förderung der wissenschaftlichen Forschung (project P17973) have funded this work.

**Figure captions**

Fig. 1.
Experimental setup. The chirped pulses generated by the $Cr^{4+}$:YAG oscillator are compressed by a single mode fiber and spectrally broadened in the PCF. The SC light probes a cell and is analyzed by a FT spectrometer. CM: chirped mirrors, OC: output coupler.

Fig. 2.
Low resolution spectra on a logarithmic vertical scale of the laser output (blue line, plot filled in white) and the 4.5 µm-core-diameter SF6 PCF output (red line, plot filled in magenta).

Fig.3.
SC high resolution spectra on a linear vertical scale of the acetylene molecule in the 1.5 µm region at 28 and 160 hPa. The strongest spectral features are the P and R branches of the $\nu_1+\nu_3$ band of $^{12}C_2H_2$.

Fig. 4.
High resolution (15 GHz) spectra (on a linear vertical scale) of the ammonia molecule in the 1.65 µm region. Lower trace: SC spectrum with a $NH_3$ pressure of 128 hPa and an absorption path length of 70 cm. Recording time: 3.1 s. Upper trace: spectrum from a classical halogen source with a $NH_3$ pressure of 132 Pa and an absorption path length of 28 m. Recording time: The vertical scale is shifted to allow better comparison. Recording time: 10 min.

**Fig. 1.**
Experimental setup. The chirped pulses generated by the $Cr^{4+}$:YAG oscillator are compressed by a single mode fiber and spectrally broadened in the PCF. The SC light probes a cell and is analyzed by a FT spectrometer. CM: chirped mirrors, OC: output coupler.

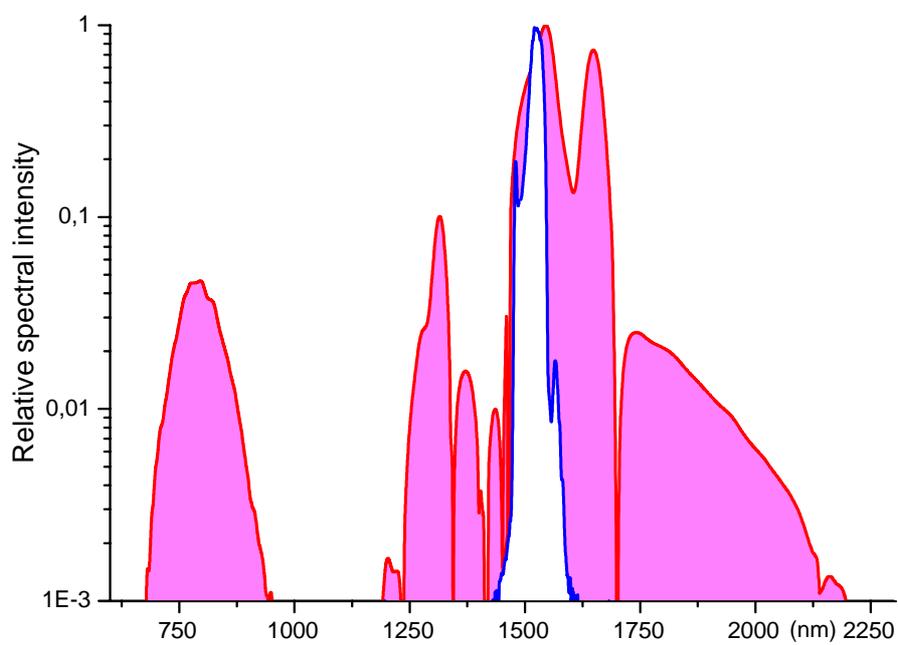

**Fig. 2.**
Low resolution spectra on a logarithmic vertical scale of the laser output (blue line, plot filled in white) and the 4.5 µm-core-diameter SF6 PCF output (red line, plot filled in magenta).

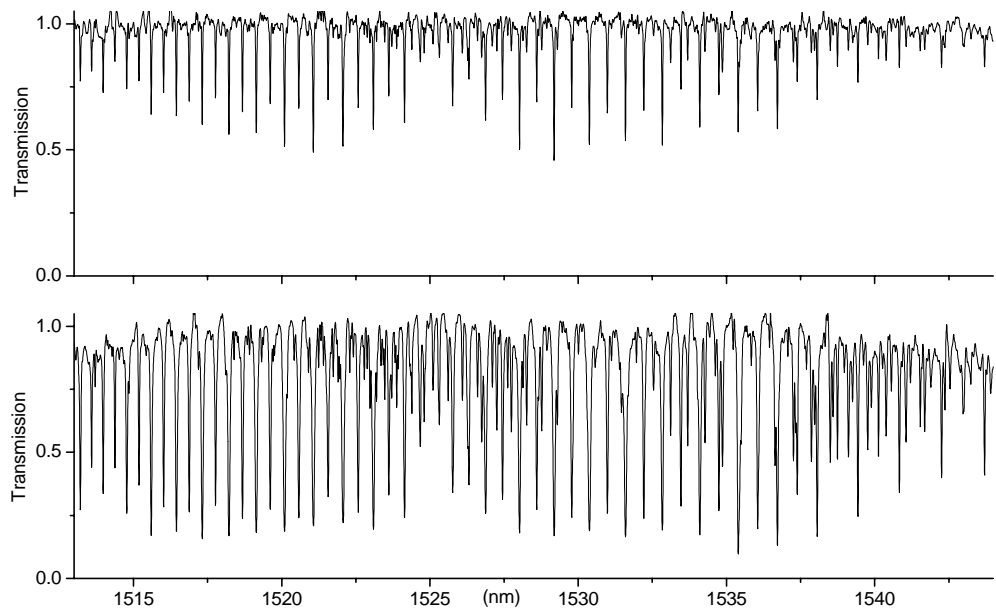

**Fig.3.**
SC high resolution spectra on a linear vertical scale of the acetylene molecule in the 1.5 µm region at 28 and 160 hPa. The strongest spectral features are the P and R branches of the $\nu_1+\nu_3$ band of $^{12}C_2H_2$.

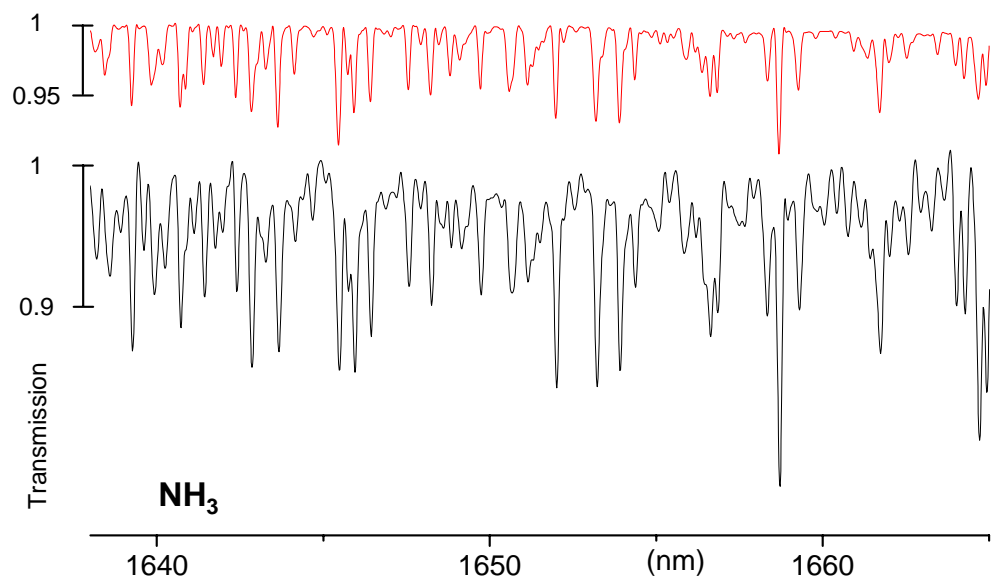

**Fig. 4.**
High resolution (15 GHz) spectra (on a linear vertical scale) of the ammonia molecule in the 1.65 µm region. Lower trace: SC spectrum with a NH$_3$ pressure of 128 hPa and an absorption path length of 70 cm. Recording time: 3.1 s. Upper trace: spectrum from a classical halogen source with a NH$_3$ pressure of 132 Pa and an absorption path length of 28 m. Recording time: The vertical scale is shifted to allow better comparison. Recording time: 10 min.

**References without titles**